\documentclass[lettersize,journal]{IEEEtran}
\usepackage{amsmath,amsfonts}
\usepackage{algorithm}
\usepackage{algpseudocode}
\usepackage{array}
\usepackage[caption=false,font=normalsize,labelfont=sf,textfont=sf]{subfig}
\usepackage{textcomp}
\usepackage{stfloats}
\usepackage{url}
\usepackage{verbatim}
\usepackage{graphicx}
\usepackage{cite}
\usepackage{xcolor}
\usepackage{amsthm}
\usepackage{amssymb}
\usepackage{amsmath,lipsum}
\usepackage{epstopdf}
\usepackage{bm}
\usepackage{makecell}
\usepackage{cuted}

\begin{document}

\title{\LARGE Semantic Communications for Digital Signals via Carrier Images}

\author{Zhigang Yan and Dong Li,~\IEEEmembership{Senior Member, IEEE }
\vspace{-6mm}
\thanks{This work was supported in part by the Science and Technology Development Fund, Macau, SAR, under Grant 0188/2023/RIA3. Zhigang Yan and Dong Li are with the School of Computer Science and Engineering, Macau University of Science and Technology, Macau, China. (e-mails: 3220005784@student.must.edu.mo; dli@must.edu.mo). (\textit{Corresponding Author: Dong Li}.)
}
}



\maketitle
\pagestyle{empty}
\thispagestyle{empty}

\begin{abstract}
Most of current semantic communication (SemCom) frameworks focus on the image transmission, which, however, do not address the problem on how to deliver digital signals without any semantic features. This paper proposes a novel SemCom approach to transmit digital signals by using the image as the carrier signal. Specifically, the proposed approach encodes the digital signal as a binary stream and maps it to mask locations on an image.  This allows binary data to be visually represented, enabling the use of existing model, pre-trained Masked Autoencoders (MAE), which are optimized for masked image reconstruction, as the SemCom encoder and decoder. Since MAE can both process and recover masked images, this approach allows for the joint transmission of digital signals and images without incurring significant communication overheads. In addition, considering the mask tokens transmission encoded by the MAE still faces extra costs, we design a sparse encoding module at the transmitter to encode the mask tokens into a sparse matrix, and it can be recovered at the receiver. Thus, this approach simply needs to transmit the latent representations of the unmasked patches and a sparse matrix, which further reduce the transmission overhead compared with the original MAE encoder. Simulation results show that the approach maintains reliable transmission even in a high mask ratio of images.
\end{abstract}
\vspace{-1mm}
\vspace{-2mm}
\section{Introduction}
\IEEEPARstart{S}{emantic} communication (SemCom), which is a promising technology in communication systems that leverages deep learning, such as neural networks, to transmit the meaning or semantics of information rather than just the raw data. This approach is particularly powerful in reducing the amount of data that needs to be transmitted and improving the efficiency, especially in scenarios with a limited bandwidth \cite{s}. SemCom is particularly well-suited for transmitting some specific types of data, such as texts \cite{text} and images \cite{image}, which inherently carry high-level semantics that can be efficiently encoded and decoded using neural network models \cite{dl}. 

Transmitting the more general type of data is essential in many wireless communication scenarios, such as offloading data in mobile edge computing \cite{mec} and model parameters in wireless federated learning \cite{fl}. However, as the volume of data and the size of models increase, this process can lead to a significant communication overhead. Although text and image transmissions are generally considered for SemCom by exploiting their rich semantic feature information \cite{text}, \cite{image}, \cite{sc2}, \cite{lstm}, digital signals for SemCom have received little attention since they typically do not contain any semantic information which renders them difficult for direct applications to SemCom. In order to fill this gap, in this paper, we design a transmission approach for binary stream of digital signals transmitted via images as carrier signals through the existing SemCom frameworks without redesigning to the joint source and channel coding. To the best of the authors' knowledge, this is the first attempt in solving the transmission of digital signals in SemCom. Our main contributions are summarized as follows:

\begin{enumerate}
\item \textbf{A novel transmission approach for joint digital signal and image via Masked Autoencoders (MAE):} We design a SemCom framework for digital signals by utilizing images for carrier signals, which first converts digital signals into a binary stream, and then transmits alongside masked images. The MAE model is integrated into the framework, which ensures a high accuracy in recovering not only digital signals and but also images.

\item \textbf{Transmission overhead reduction by the sparse encoding module:} Different from the original MAE model with a fixed length of mask tokens, we consider a sparse encoding module to further encode the mask tokens output from the MAE encoder into a sparse matrix, to reduce transmission costs. This sparse matrix is recovered at the decoder on the receiver side with the latent representations of the unmasked patches to reconstruct both digital signals and images.

\item \textbf{Robustness in image variations:} Simulation results demonstrate the robustness of the proposed approach in digital signal transmission even under a high mask ratio.
\end{enumerate}

\vspace{-3mm}
\section{System Model}

In this section, we first introduce the SemCom framework and the pre-trained MAE model. We then design a novel transmission approach based on this MAE-based SemCom framework, enabling the simultaneous transmission of digital signals along with images.
\vspace{-4mm}
\subsection{SemCom Framework}

The SemCom framework under consideration consists of a semantic encoder/decoder and a channel encoder/decoder. Specifically, at the transmitter, the source data $\mathbf{S}$ is first encoded into a symbol stream $\mathbf{X}$ through the semantic encoder and channel encoder. Let $f_{1}(\cdot)$ represent the semantic encoding function and $f_{2}(\cdot)$ the channel encoding function. The output of the transmitter can be expressed as
\begin{equation}\small
	\label{trans}
	\mathbf{X} = f_{2}(f_{1}(\mathbf{S})),
\end{equation}
where $\mathbf{X}\in\mathbb{C}^{N_{t}\times 1}$ and $N_{t}$ is the number of antennas at the transmitter. The encoded signal $\mathbf{X}$ is then transmitted over the physical channel to the receiver. At the receiver, the received signal $\mathbf{Y}$ is given by
\vspace{-2mm}
\begin{equation}\small
	\label{xy}
	\mathbf{Y} = \mathbf{H}\mathbf{X}+\mathbf{N},
\end{equation}
where $\mathbf{H}\in\mathbb{C}^{N_{r}\times N_{t}}$ is the channel matrix, $\mathbf{N}\in\mathbb{C}^{N_{r}\times 1}$ represents additive white Gaussian noise (AWGN), and $N_{r}$ is the number of receiver antennas. The noise follows $\mathbf{N} \sim \mathcal{CN}(0, \sigma^{2}\mathbf{I})$. Finally, the receiver recovers the original source data $\mathbf{S'}$ through the channel decoder and semantic decoder. This can be expressed as
\begin{equation}\small
	\label{rev}
	\mathbf{S'} = f_{1}^{-1}(f_{2}^{-1}(\mathbf{Y})).
\end{equation}
The semantic encoder and decoder are usually trained together in an offline setting, using a large dataset, often referred to as the shared knowledge. The goal during training is to minimize the difference between $\mathbf{S}$ and $\mathbf{S'}$, ensuring that the semantic information encoded by the transmitter is accurately decoded by the receiver. This process involves jointly optimizing the transmitter and receiver models. Once trained, the models are deployed into the transmitter and the receiver for real-time semantic encoding and decoding during communication.

\vspace{-4mm}
\subsection{Problem Description}

As we mentioned before, current SemCom frameworks can offer excellent performance on the image transmission, which, however, do not address how to transmit digital signals. In this regard, we propose and investigate a novel SemCom framework to transmit digital signal by the aid of images, which does not require to redesign existing SemCom encoder/decoder structures. Specifically, the source data $\mathbf{S}$ in this framework includes the image $\mathbf{S}_{\rm im}$ and the digital signal $\mathbf{S}_{\rm ds}$, and the goal of SemCom model training is minimizing the gap between $\mathbf{S}_{\rm im}$, $\mathbf{S}_{\rm ds}$ and the reconstruction results $\mathbf{S}_{\rm im}'$, $\mathbf{S}_{\rm ds}'$ with affordable transmission overhead, which can be formulated as a multiple objectives optimization problem with the same constraint, and is given by
\begin{align}
	&\mathbf{P}:~\min_{\mathbf{w}}~ \Vert\mathbf{S}_{\rm im}-\mathbf{S}_{\rm im}'\Vert_{2}^{2}\tag{P1a}  \label{P1}\\
	&~~~\quad\min_{\mathbf{w}}~ \Vert\mathbf{S}_{\rm ds}-\mathbf{S}_{\rm ds}'\Vert_{2}^{2}\tag{P1b}  \label{P2}\\
	&~~~\quad\;\textrm{s.t.}\quad {\rm Cost}(\mathbf{X})\leq \epsilon, \tag{C1} \label{P1a}
\end{align}
where $\mathbf{X}$ is the parameter of the semantic encoder/decoder, ${\rm Cost}(\cdot)$ denotes the number of bits required for transmission and $\epsilon$ is its threshold. Note that, different from traditional communication systems where the carrier signal contains no information, the images serving as the carrier signal in our proposed SemCom has its own information to deliver.

\vspace{-4mm}
\section{Proposed Transmission Approach}

This section focuses on the SemCom framework design for the joint digital signal and image transmission. 
\vspace{-4mm}
\subsection{An MAE-based Backbone SemCom Framework}

The MAE consists of an encoder that maps the input image to a latent representation and a decoder that reconstructs the image from this latent space. Furthermore, MAE's distinct asymmetric design allows it to reconstruct the original image even when some input patches are masked. Its structure is illustrated in Fig. 1(a).  

Notably, the MAE model is robust to varying distributions of masked patch locations \cite{mae}. This characteristic allows us to map raw data into a specific masking strategy and then mask corresponding patches of an image. Upon receiving the masked image, the transmitter with MAE encoder processes the visible patches and identifies the indices of the masked ones and transmit the results to the receiver over the wireless channel. The decoder on the receiver then reconstructs the original image. Finally, the receiver reverses the index sequence of the masked patches to recover the original data. Thanks to the exceptional image reconstruction capabilities of the MAE, this SemCom framework excels not only in traditional image transmission tasks, but also enables the transmission of digital signals that were challenging to handle in existing SemCom frameworks. This is achieved without incurring additional communication costs.

The transmission process is summarized in Fig. 1(b). As shown in this figure, the MAE-based SemCom framework can not only reconstruct the image from the received semantic information by minimizing \eqref{P1}, but also recover digital signals from the mask tokens by minimizing \eqref{P2} without the additional transmission costs. Specifically, let $\mathbf{S}_{\rm im} \in \mathbb{R}^{C \times H \times W}$ denote the input image, where $C$ is the number of channels, and $H$ and $W$ are the height and width of the image, respectively. The image can be divided into patches, represented as $\{s_{1}, \cdots, s_{N}\}$, where $s_{n} \in \mathbb{R}^{C \times P \times P}$, $N$ is the total number of patches with $P \times P$ being the size of each patch, and these patches are indexed from $1$ to $N$. Moreover, the masked patch locations are determined by the indices of ``1"s in the binary sequence. For each ``1'', a corresponding image patch is masked, while the ``1''s represent unmasked patches. This binary-to-mask mapping allows for the efficient transmission of arbitrary data types as masked images. In summary, if the indices of the masked patches are represented by the set $\mathcal{M}$, the input to the encoder consists only of the visible patches, denoted as $\mathbf{S}_{\rm im}^{v} = \{s_{n} \mid n \notin \mathcal{M}\}$, and the indices of the missing patches. Moreover, the outputs of the encoder include the latent representation of $\mathbf{S}_{\rm im}^{v}$ and the mask tokens $\mathbf{M}$. Then, they are sent to the receiver over the wireless channel and decoded to the reconstructed image $\mathbf{S}_{\rm im}'$ and the received digital signal $\mathbf{S}_{\rm ds}'$, which can be written as
\begin{equation}
	\label{mae}
	\{\mathbf{S}_{\rm im}', \mathbf{S}_{\rm ds}'\} = g_{\rm de}(g_{\rm en}(\mathbf{S}_{\rm im}^{v}),\mathbf{M}),
\end{equation}
where $g_{\rm en}(\cdot)$ and $g_{\rm de}(\cdot)$ represent the encoder and decoder functions, respectively. In particular, the digital bits can be recovered by detecting the mask locations in the masked images. Since the MAE model is already proficient in compressing, encoding, and decoding masked images, we can leverage this well-established framework without needing to redesign the entire SemCom system. Compared with current SemCom frameworks for image transmission, such as \cite{image,sc2,jscc,swim}, which encode a full image, the proposed MAE-based framework only encodes the unmasked patches and transmits binary data by the mask locations. Thus, it not only reduces the overhead of image transmission, but also transmits the digital signal without extra costs.
\vspace{-3mm}
\subsection{Proposed Transmission Approach}

\begin{figure}[!t]
	\centering
	\subfloat[]{\includegraphics[width=3.2in]{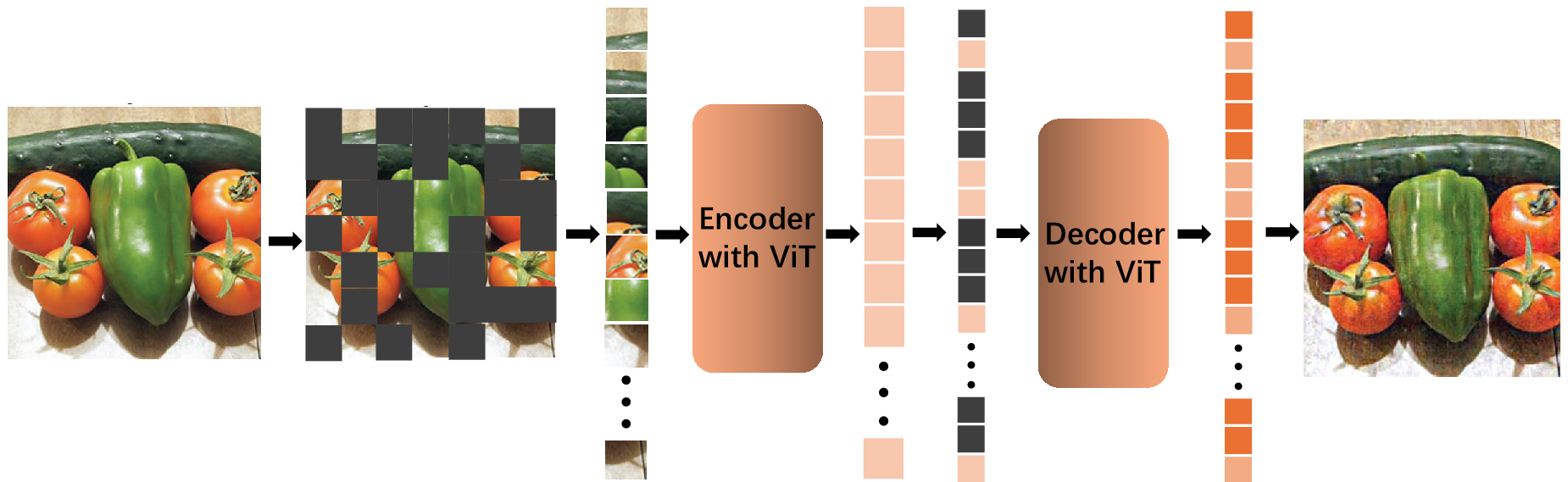}
		\label{fig2_1}}\\
	\vspace{-2mm}
	\subfloat[]{\includegraphics[width=3.2in]{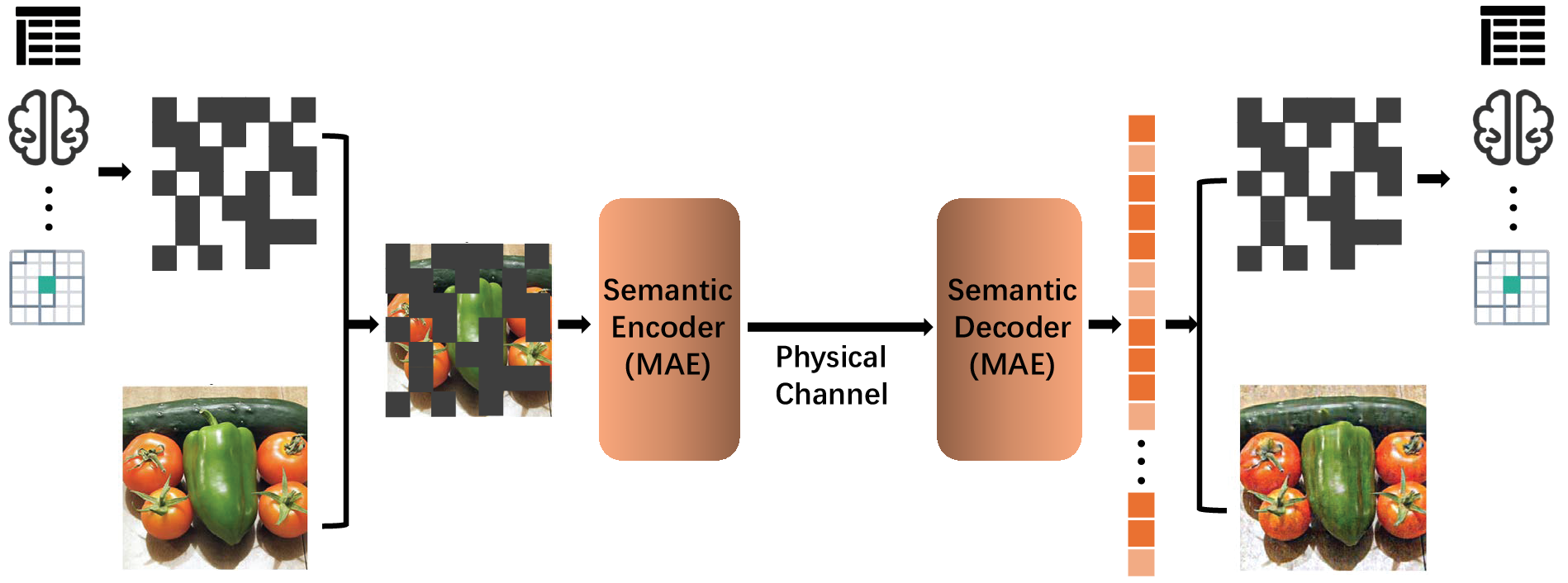}
		\label{fig2_2}}\\
	\vspace{-2mm}
	\subfloat[]{\includegraphics[width=3.2in]{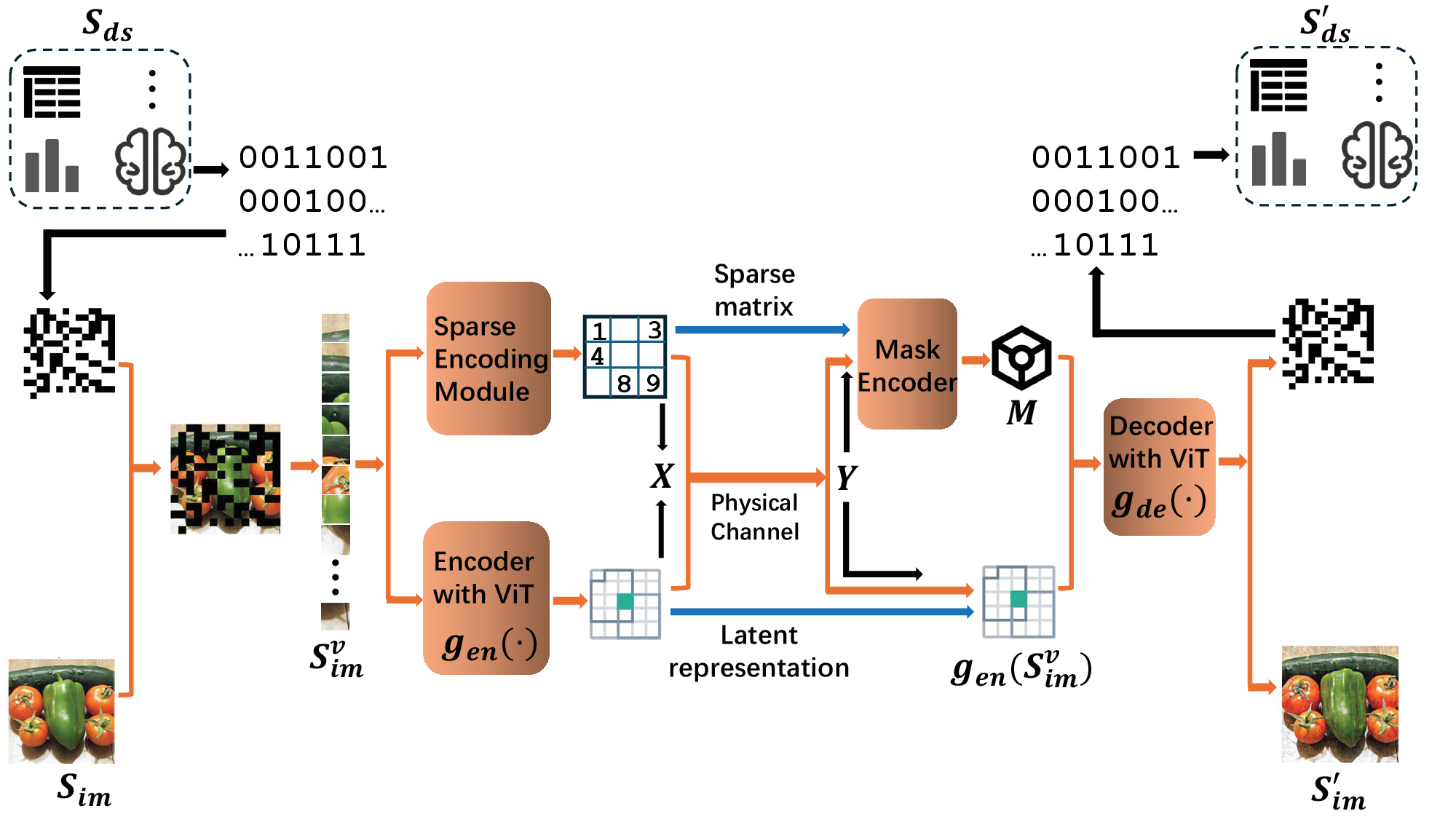}
		\label{fig2_3}}
	\caption{(a) The MAE architecture. (b) The MAE-based SemCom framework. (c) The proposed transmission approach.}
\end{figure}

In the proposed SemCom transmission framework, 
we apply the MAE encoder/decoder as the SemCom encoder/decoder. The outputs of the original MAE encoder include the latent representation of the unmasked patches of the image $g_{\rm en}(\mathbf{S}_{\rm im}^{v})$ and the mask tokens $\mathbf{M}$. While the original MAE efficiently transmits the image's most important features by sending $g_{\rm en}(\mathbf{S}_{\rm im}^{v})$, it still carries a significant overhead in transmitting $\mathbf{M}$. Since the size of $\mathbf{M}$ is constant, regardless of how many patches are masked, we will have unnecessary transmission of information, and inefficient utilization of the available resources that may result in high transmission costs. To address these challenges and further optimize the MAE-based SemCom framework, we add a sparse encoding module at the transmitter to encode the mask tokens into a sparse matrix to reduce the transmission costs, and this sparse matrix can be recovered to the tokens at the receiver. Thus, both the recovered mask tokens and received latent representation are the inputs to the MAE decoder to obtain the transmitted image and digital signal. The whole proposed transmission approach is shown in Fig. 1(c).

\begin{algorithm}[!t]
	\caption{Sparse Encoding Module}
	\begin{algorithmic}[1]
		\State \textbf{Input:} $g_{\rm en}(\mathbf{S}_{\rm im}^{v})$, $\mathbf{M}$
		
		\State $L \gets \text{Length of }\mathbf{M}$ \Comment{Total number of patches}
		\State $\text{len\_keep} \gets\text{Length of }g_{\rm en}(\mathbf{S}_{\rm im}^{v}) - 1$ \Comment{Number of unmasked patches}
		
		\State $\text{ids\_0} \gets \text{ids\_restore}[:, :\text{len\_keep}]$
		\Comment{indices of ``0''}
		\State $\text{ids\_1} \gets \text{ids\_restore} \setminus \text{ids\_restore}[:, :\text{len\_keep}]$
		\Comment{indices of ``1''}
		\State $\text{full\_matrix} \gets \{0, 1, \dots, L-1\}$
		\Comment{Set of all possible patch indices}
		\If{\text{length of ids\_0} $>$ \text{length of ids\_1}}
		\State $\text{sparse\_matrix} \gets \text{full\_matrix} \setminus \text{ids\_0}$
		\Else
		\State $\text{sparse\_matrix} \gets \text{full\_matrix} \setminus \text{ids\_1}$
		\EndIf
		\State \textbf{Output:} \text{sparse\_matrix}
	\end{algorithmic}
\end{algorithm}

\begin{algorithm}[!t]
	\caption{Encoding sparse matrix for mask tokens}
	\begin{algorithmic}[1]
		\State \textbf{Input:} Received $g_{\rm en}(\mathbf{S}_{\rm im}^{v})$, \text{sparse\_matrix}
		\State $L_{m} \gets \text{Length of sparse\_matrix}$ \Comment{Number of masked patches}
		\State $L \gets L_{m}+\text{Length of }g_{\rm en}(\mathbf{S}_{\rm im}^{v}) - 1$ \Comment{Total number of patches}
		\State $\text{mask}\gets\mathbf{0}\in\mathbb{R}^{3\times L}$
		\State $\text{mask}\gets\text{mask}[:,\text{sparse\_matrix}]=1$ \Comment{Generate the full matrix}
		\State \text{ids\_shuffle} $\gets$ \text{argsort(mask, dim $=1$)}  \Comment{Sort the mask along the length dimension}
		\State $\mathbf{M}\gets$ \text{argsort(ids\_shuffle, dim $=1$)} \Comment{Get the inverse sort indices to restore the original order}
		\State \textbf{Output:} Recovered mask tokens $\mathbf{M}$
	\end{algorithmic}
\end{algorithm}

In this proposed approach, the transmitted data $\mathbf{S}_{\rm ds}$ is encoded to binary bits firstly, which is mapped to a mask sampling strategy and applied to the transmitted image $\mathbf{S}_{\rm im}$. Then, this masked image is inputted to the proposed encoder, which includes an MAE encoder and a sparse encoding module, which is shown in Algorithm 1. The MAE encoder outputs the latent representation of this image (e.g., $g_{\rm en}(\mathbf{S}_{\rm im}^{v})$) and the mask tokens from the masked patches and the masked indices (e.g., $\mathbf{M}$). Next, the obtained mask tokens are encoded to a sparse matrix, and both this matrix and the latent representation are sent to the receiver from the transmitter over wireless channels. After receiving these two parts, the receiver inputs the received sparse matrix into a mask encoder to recover the mask tokens. This process is shown in Algorithm 2. Then, these tokens and the received latent representation are sent to the MAE decoder, and the reconstructed image $\mathbf{S}_{\rm im}'$ and recovered binary bits are obtained. Finally, the transmitted data $\mathbf{S}_{\rm ds}'$ is decoded from the binary data.

\vspace{-4mm}
\subsection{Transmission Overhead Analysis}

The number of transmission bits of this approach is depended on the structure of the MAE model, the size of the image and the length of the binary data for transmission. Note that, the mask ratio of the raw image becomes larger for a longer binary data, which makes fewer patches of the image needed to be encoded by the MAE. Thus, it reduces the number of transmitted bits compared to encoding the entire image. Specifically, if $L_{e}$ and $L_{m}$ denote the required number of bits for transmitting $g_{\rm en}(\mathbf{S}_{\rm im}^{v})$ and $\mathbf{M}$, respectively, the number of transmitted bits for the proposed approach is given by $\lceil \frac{W\times H}{P\times P} \rceil\times(1-M_{r})L_{e}+L_{m}$, where $M_{r}$ is the masked ratio. Moreover, $L_{e}$ and $L_{m}$ are determined by the MAE structure and the transmitted binary stream. Furthermore, if the length of the binary data is $L_{b}$, we have $M_{r}\leq L_{b}\times \lceil \frac{W\times H}{P\times P} \rceil^{-1}$. Thus, the minimal number of transmitted bits with a fixed masked ratio can be written as $(\lceil \frac{W\times H}{P\times P} \rceil-L_{b})L_{e}+L_{m}$. Note that, this value is decreasing with a larger $L_{b}$, so adding the transmitted bits of the binary data can reduce the transmission overhead of the proposed approach. In addition, since the pre-trained MAE model performs well in a wide range of mask ratios (about 10-80$\%$) \cite{mae}, the performance on the image transmission keeps stable with $L_{b}$ increasing in this range. On the contrary, transmitting the image and binary data respectively (without SemCom) requires $W\times H\times C \times B_{p}+L_{b}$ bits, where $C$ is the number of channels and $B_{p}$ is the required number of bits for each pixel.

\begin{figure}[!t]
	\vspace{-12mm}
	\centering
	\subfloat[]{\includegraphics[width=1.7in]{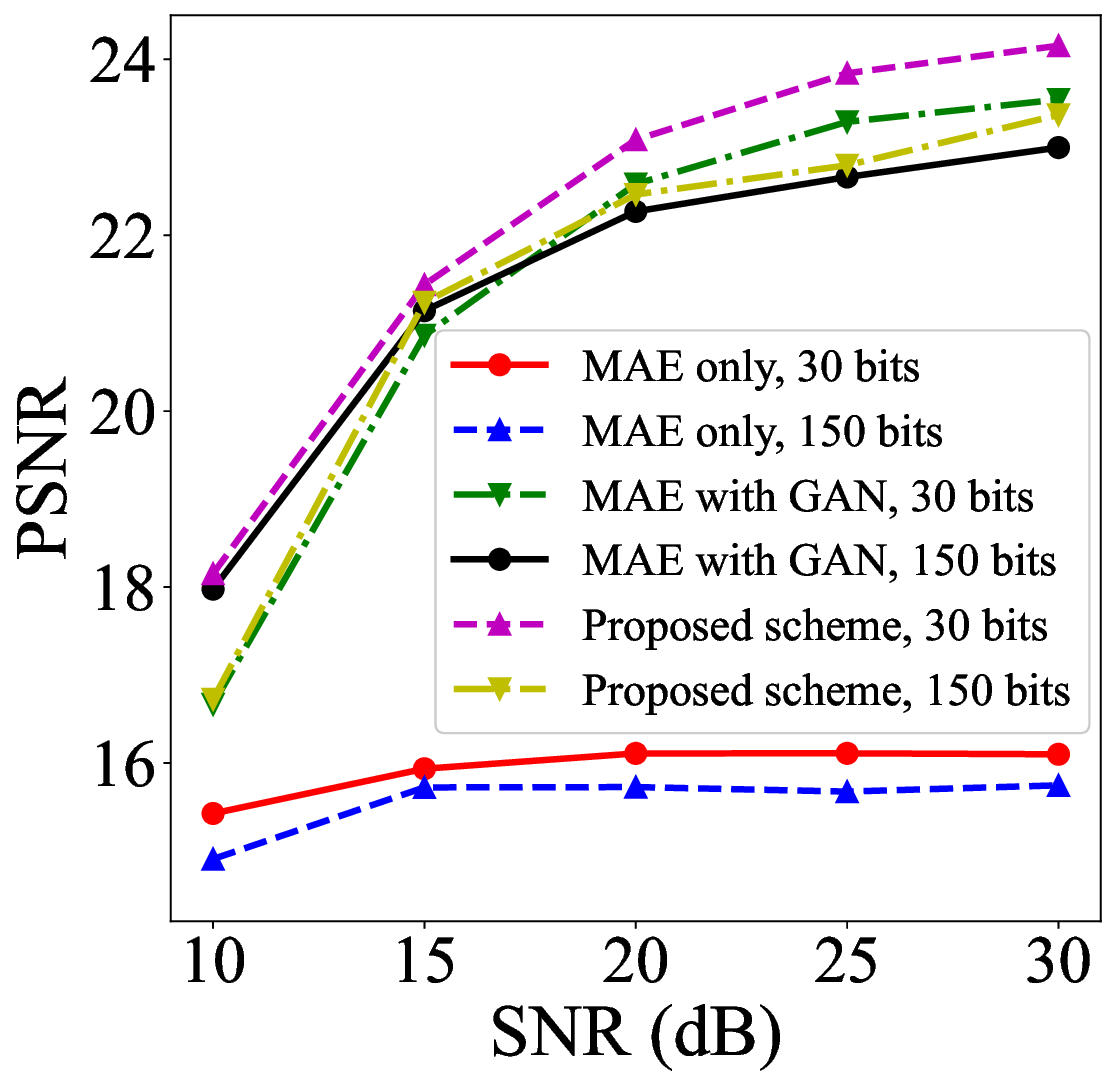}
		\label{fig5_1}}
	\subfloat[]{\includegraphics[width=1.7in]{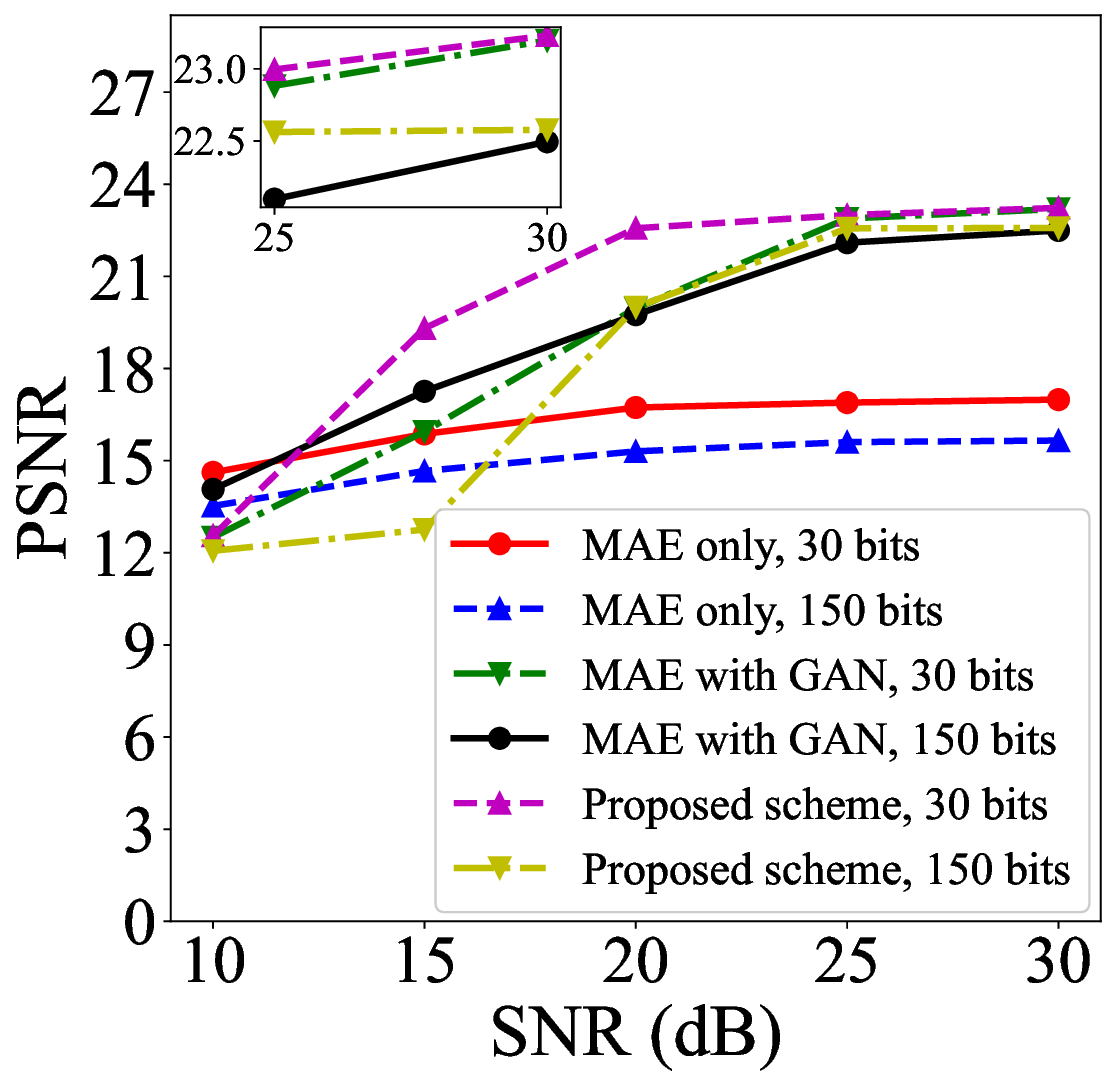}
		\label{fig5_2}}
	\caption{The PSNR performance of image reconstruction under different SNRs. (a) AWGN channel. (b) Rayleigh fading channel.}
\end{figure}
\vspace{-5mm}
\section{Simulation Results}

In this section, we present the simulation setup and results, which illustrate the effects of varying channel conditions and demonstrate the effectiveness of our proposed SemCom framework for both image and digital signal transmissions.
\vspace{-4mm}
\subsection{Simulation Setup}

The transmitted images are selected from the validation set of ImageNet-1K and CIFAR-10, comprising the size of $224\times 224$ and $32\times 32$ images, while transmitted binary signals are randomly generated with varying lengths. For image transmission evaluation, Peak Signal to Noise Ratio (PSNR) and Multi-Scale Structural Similarity (MS-SSIM) are used as metrics to assess the pixel-level quality and structural similarity, respectively. Binary transmission performance is evaluated by the bit error rate (BER). Simulations are conducted over both AWGN and Rayleigh fading channels.

Unlike prior SemCom research which primarily focused on enhancing the image reconstruction quality, our work aims to establish a SemCom-based transmission approach for both digital signals and images. Thus, the simulations are organized as follows: (a) Verify the feasibility and evaluate the performance of the proposed approach for both image and digital signal transmission, and (b) Compare communication overhead with existing SemCom frameworks. Specifically, we assess the proposed approach’s performance and compare the bit requirements for transmitting one image and binary data of varying lengths against previous SemCom models, including DeepJSCC-V \cite{jscc} and SwinJSCC \cite{swim}, which are CNN and ViT-based models optimized for image transmission and require additional overhead for binary data.

In addition, two loss functions are designed for MAE training in previous works, which are the mean squared error (MSE) \cite{mae} and a combined loss with MSE and GAN loss \cite{gan}, written as \textit{MAE only} and \textit{MAE with GAN} in this paper. \textit{MAE only} is primarily designed for self-supervised learning and representation learning, rather than specifically for image reconstruction \cite{mae}. To improve the realism of the reconstructed images, \textit{MAE with GAN} is proposed in \cite{gan}. Thus, \textit{MAE with GAN} is considered in our framework, and \textit{MAE only} is used as a benchmark.
\vspace{-5mm}
\subsection{Simulation Results}

\begin{figure}[!t]
	\vspace{-1cm}
	\centering
	\subfloat[]{\includegraphics[width=1.7in]{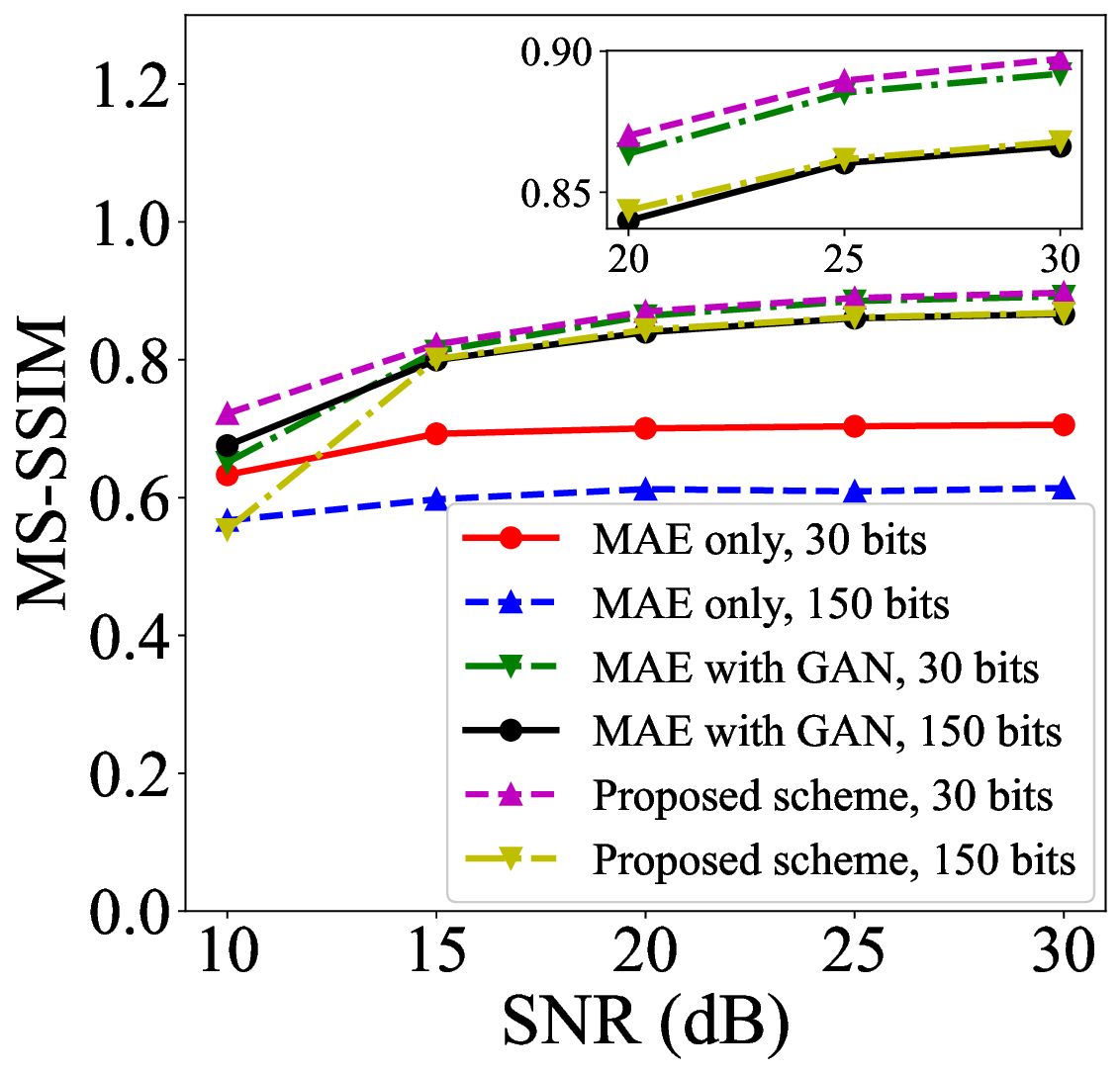}
		\label{fig6_1}}
	\subfloat[]{\includegraphics[width=1.7in]{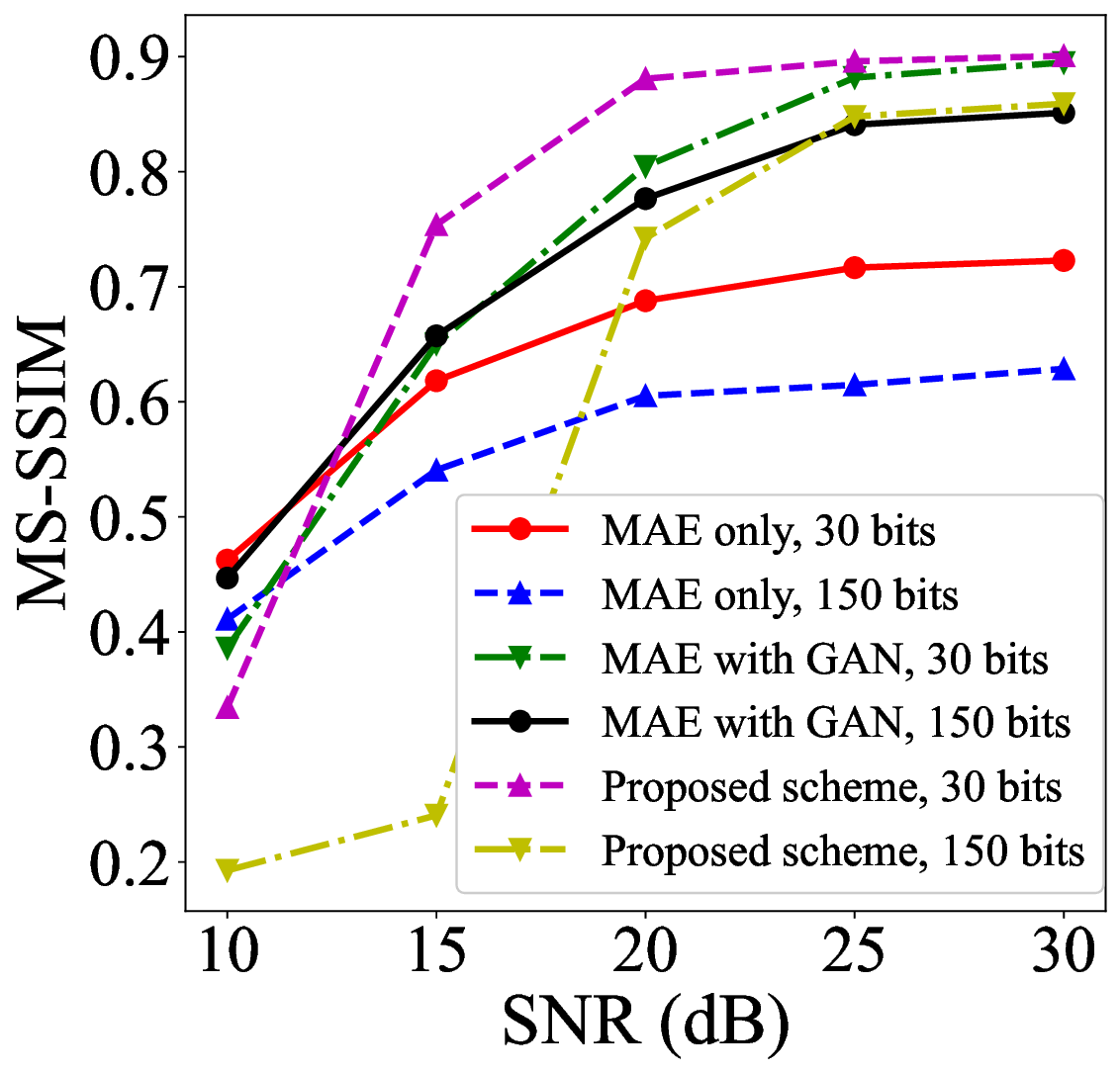}
		\label{fig6_2}}
	\caption{The MS-SSIM performance of image reconstruction under different SNRs. (a) AWGN channel. (b) Rayleigh fading channel.}
\end{figure}
\textit{a) Performance of the image transmission:} To verify the effectiveness of our proposed approach for image transmission, we compare PSNR and MS-SSIM results for the proposed approach and two MAE-based SemCom frameworks without sparse module and trained on \textit{MAE only} and \textit{MAE with GAN}, highlighting the benefits of our approach design.

Fig. 2 presents the PSNR performance across varying SNRs in both AWGN and Rayleigh fading channels. In low-SNR conditions, the \textit{MAE only} approach yields a lower PSNR, regardless of the sparse module application. Moreover, as the SNR increases, the proposed approach shows significant PSNR improvement over the \textit{MAE with GAN} approach without the sparse module. Additionally, the proposed approach not only achieves a higher PSNR in high-SNR scenarios but also exhibits enhanced robustness to changes in the transmitted binary data length. These performance advantages in image transmission are also highlighted by the MS-SSIM results.

Fig. 3 presents MS-SSIM performance versus SNR for AWGN and Rayleigh fading channels. Unlike PSNR, which focuses on the pixel similarity, MS-SSIM evaluates the perceptual quality, aligning more closely with the human visual perception. In Fig. 3, the proposed approach shows improvement over benchmarks as SNR increases. These results highlight the importance of integrating sparse encoding module into MAE to enhance the image reconstruction quality of SemCom. In addition, all these schemes perform well regarding the image recovery when fewer bits are transmitted.

\textit{b) Performance of the binary data transmission:} Fig. 4 illustrates the BER for binary data transmission using the proposed approach across different SNRs and numbers of masked patches over AWGN channels. In Fig. 4(a), as SNR increases, the BER performance improves. Fig. 4(b) shows that, compared with the \textit{MAE with GAN}, the proposed approach performs well regarding the BER. This result indicates that transmitting the sparse matrix of mask tokens is more resilient over wireless channels compared to transmitting mask tokens directly. Moreover, although we can observe a performance trade-off between image and digital signal transmission by increasing the transmission bits in Figs. 2, 3 and 4, this impact is not significant, since the MAE has stable feature extraction capabilities in a wide range of mask ratios \cite{mae}.

\textit{c) Performance in interference scenarios:} In addition to the simple scenarios with AWGN and Rayleigh fading, we also evaluate the performance of the proposed approach on image and digital signal transmission in interference scenarios with AWGN channel and multipath Rayleigh fading, where the number of path is 3 and the interference power is 20dBm. The results are shown in Fig. 5. Specifically, Fig. 5(a) presents the impact of the mask ratio on the performance of image transmission. A larger mask ratio leads to a worse image recovery result at the receiver. Moreover, the BER of the digital transmission on the cases of different image sizes and different mask sizes are compared in Fig. 5(b). The results reveal that with the same mask size, using a small size carrier image can reduce the BER, but using a smaller mask can improve the accuracy with the same carrier image size.

\textit{d) Transmission overhead comparison:} Table I presents the expected number of transmission bits required for a single image and associated binary data in our proposed approach compared to three baseline methods with the same size of the transmitted latent representations. Since the DeepJSCC-V and SwimJSCC only support the image transmission, their overhead includes both the compressed image and direct binary data transmission, increasing with the amount of binary data. By contrast, the proposed approach and the MAE model without the sparse encoding module leverage higher mask ratios as binary data increases, reducing the number of encoded patches and thus the transmission overhead. Additionally, by encoding only the unmasked patches and sparsely the transmitted mask tokens, our approach achieves a higher compression ratio.

\begin{figure}[!t]
	\vspace{-12mm}
	\centering
	\subfloat[]{\includegraphics[width=1.7in]{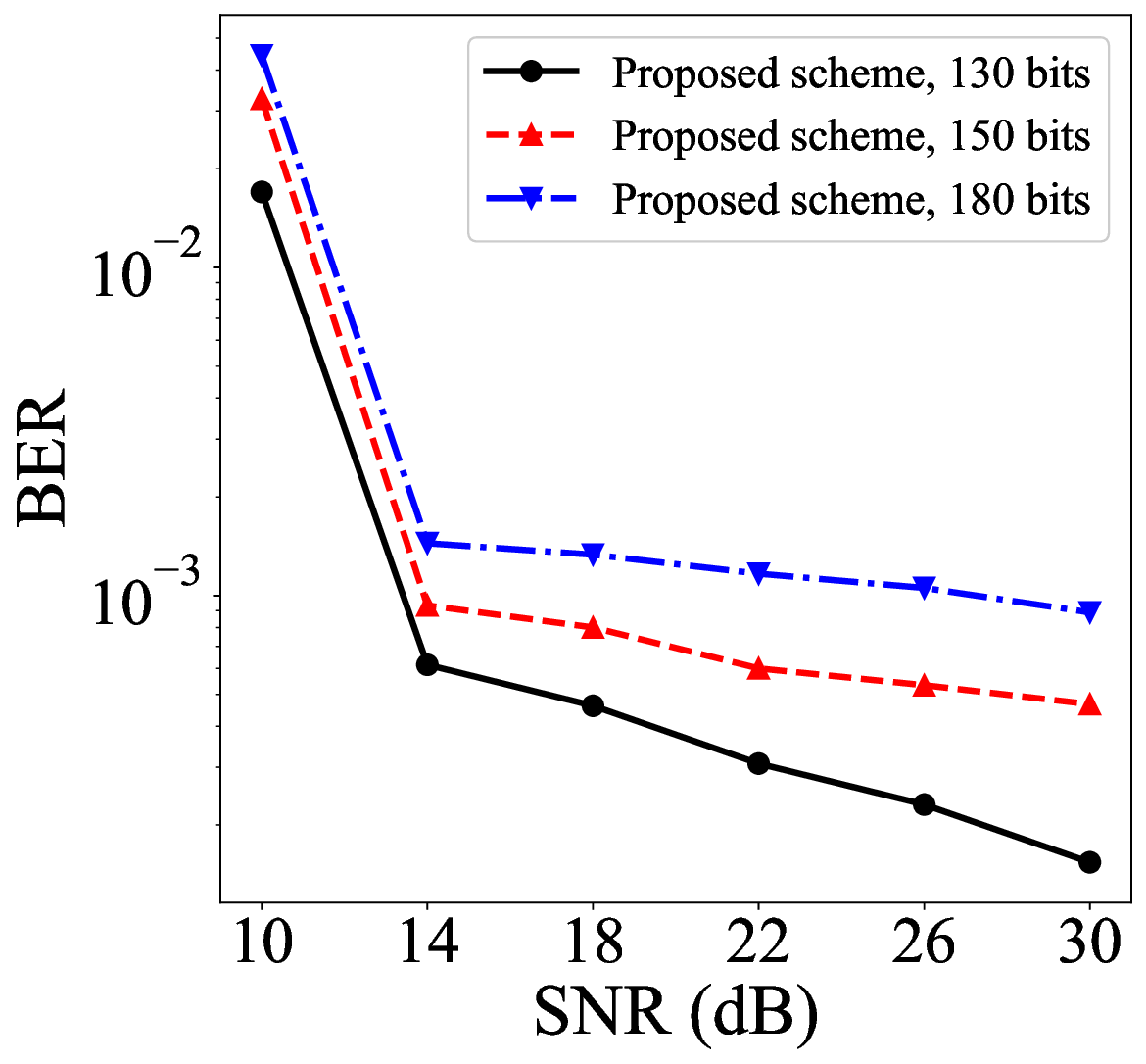}
		\label{fig7_1}}
	\subfloat[]{\includegraphics[width=1.7in]{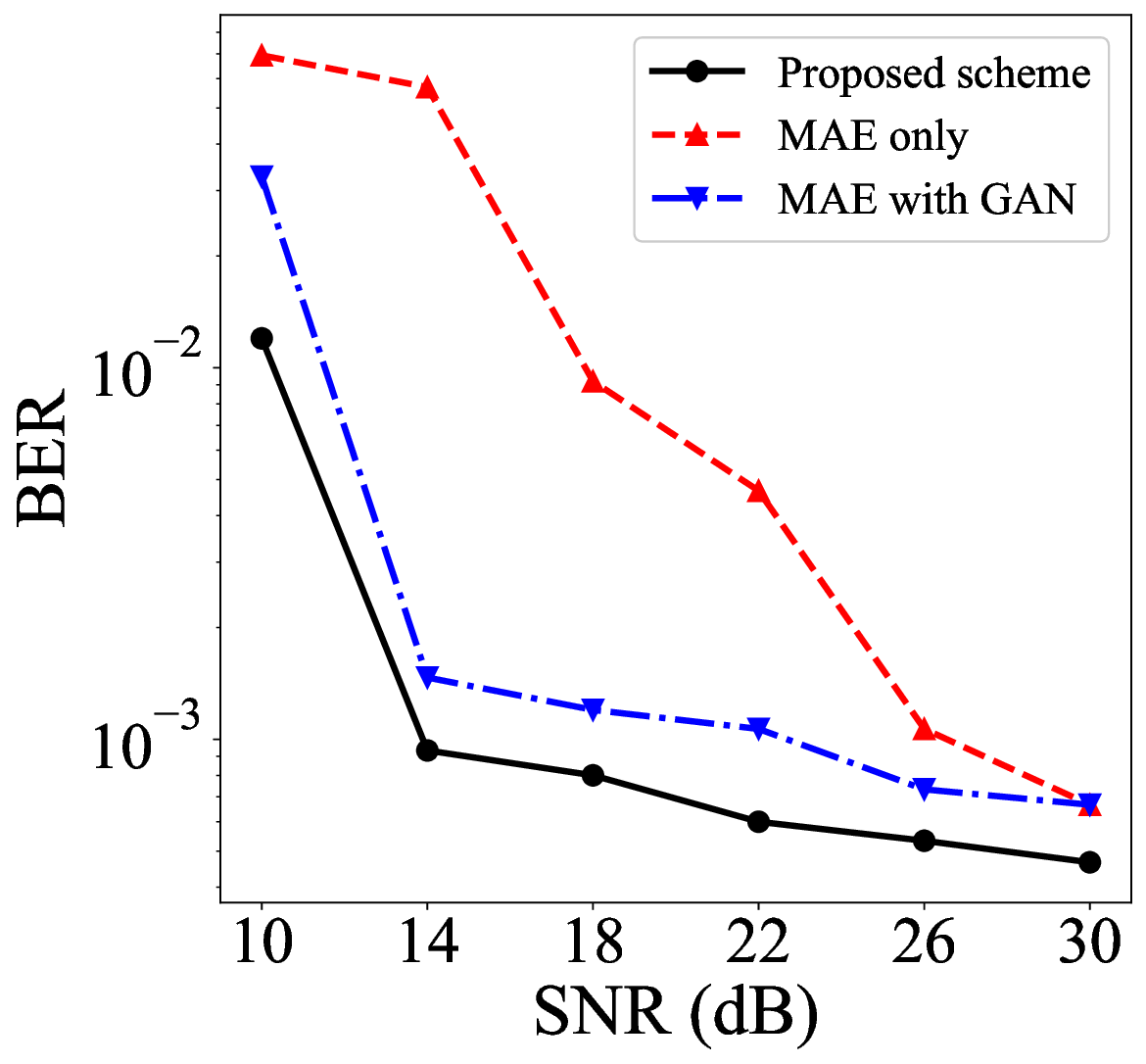}
		\label{fig7_2}}
	\caption{The BER performance of binary data transmission under different SNR. (a) BER v.s. SNR with different transmitted bits. (b) BER v.s. SNR with different schemes.}
\end{figure}

\textit{e) Computation costs comparison:} Table II provides the comparison of the time costs of the MAE with/without the proposed sparse encoding module on different sizes of images and different lengths of digital signals. As is shown, encoding a larger image with more unmasked patches costs more time, and applying the sparse encoding takes extra time costs. However, the total time is acceptable, and the increase in the calculation time is mainly caused by the increase in the image size and the decrease in the mask ratio. This is because compared with the sparse coding of the digital signal, learning-based image coding takes more time.

\vspace{-3mm}
\section{Conclusions}

In this paper, we propose a novel transmission approach that extends the capabilities of SemCom frameworks to handle more diverse and complex digital signals without semantics by leveraging the MAE model. By mapping the binary data to the locations of masked patches in an image, this approach enables joint transmission of digital signal and image without additional communication overhead. The pre-trained MAE model, optimized for handling masked image semantics, is seamlessly integrated into the system to reconstruct both the image and the corresponding binary data. Then, we design a sparse encoding module on MAE encoder to further reduce the costs of the mask tokens transmission. The simulation results validate the robustness and efficiency of the proposed approach under various channel conditions.

\begin{figure}[!t]
	\vspace{-12mm}
	\centering
	\subfloat[]{\includegraphics[width=1.7in]{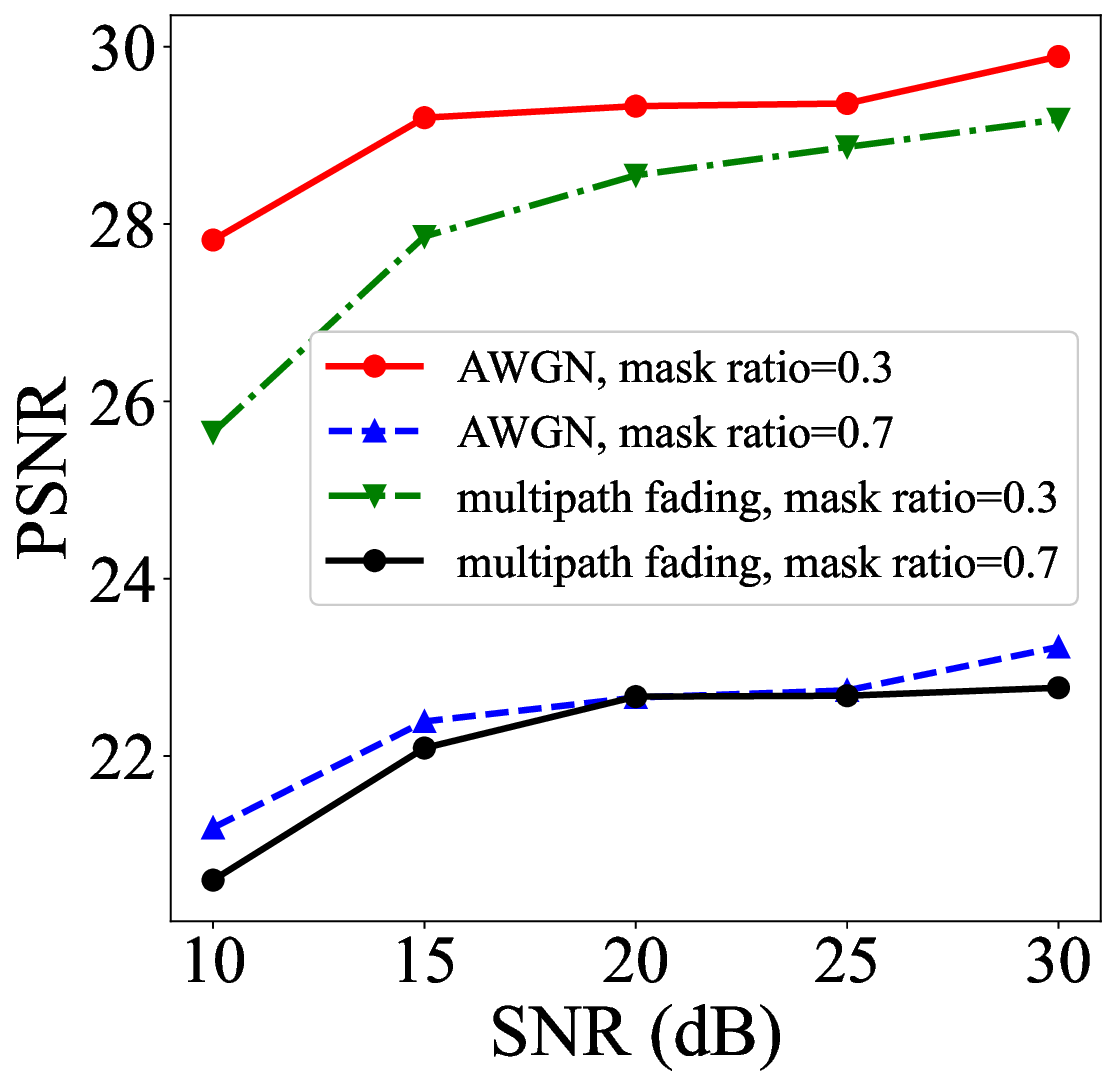}
		\label{fig8_1}}
	\subfloat[]{\includegraphics[width=1.7in]{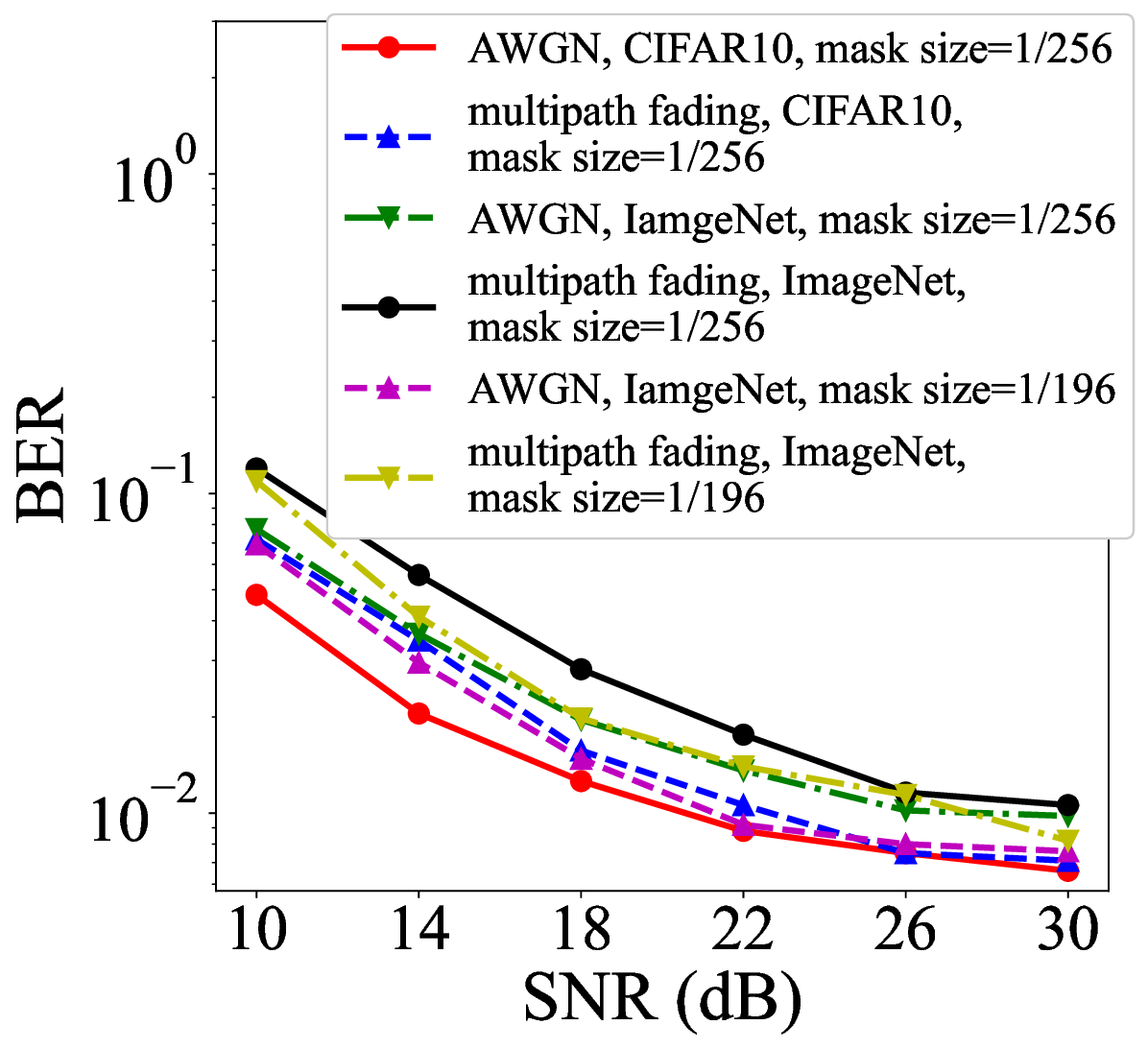}
		\label{fig8_2}}
	\caption{The PSNR and BER performance of image and binary data transmission under different SNR. (a) PSNR v.s. SNR. (b) BER v.s. SNR.}
\end{figure}

\begin{table}[!t]\scriptsize
	\vspace{-4mm}
	\caption{The Number of Transmission Bits for One Image and Binary Data of Some Specific Lengths ($\times 10^5$ bits)\label{table0}}
	\centering
	\begin{tabular}{c|ccc}
		\Xhline{1pt}
		&\makecell{One image\\$\&$100 bits} & \makecell{One image\\$\&$120 bits} & \makecell{One image\\$\&$150 bits}\\
		\hline
		\makecell{Transmit directly} & $24.0848$ & $24.0849$ & $24.0860$\\
		\hline
		\makecell{DeepJSCC-V \cite{jscc}} & $2.0080 (8.34\%)$ & $2.0082 (8.34\%)$ & $2.0086 (8.34\%)$\\
		\hline
		\makecell{SwimJSCC \cite{swim}} & $1.5063 (6.25\%)$ & $1.5065 (6.25\%)$ & $1.5068 (6.26\%)$\\
		\hline
		\makecell{Traditional MAE} & $1.5578 (6.47\%)$ & $1.4554  (6.04\%)$ & $1.3018  (5.40\%)$\\
		\hline
		\makecell{Proposed Approach} & $1.5030 (6.24\%)$ & $1.4022  (5.82\%)$ & $1.0998  (5.19\%)$\\
		\Xhline{1pt}
	\end{tabular}
\end{table}

\begin{table}[!t]\scriptsize
	\vspace{-4mm}
	\caption{The Calculation Time of the MAE with/without Sparse Encoding Module \label{table1}}
	\centering
	\begin{tabular}{c|ccc}
		\Xhline{1pt}
		&\makecell{50 bits} & \makecell{100 bits} & \makecell{150 bits}\\
		\hline
		\makecell{$32\times 32$ image} & 0.0563s/0.0545s & 0.0533s/0.0522s & 0.0505s/0.0492s\\
		\hline
		\makecell{$224\times 224$ image} & 0.7459s/0.7402s & 0.6650s/0.6549s & 0.5984s/0.5798s\\
		\Xhline{1pt}
	\end{tabular}
\end{table}

\vspace{-4mm}
\bibliographystyle{ieeetr}
\bibliography{ref1}

\vfill

\end{document}